# Experimental and Numerical Studies of the Collapse of Dense Clouds Induced by Herbig–Haro Stellar Jets

Marin Fontaine[1], Clotilde Busschaert[1], Yaniss Benkadoum[2], Isabeau A. Bertrix[2,3], Michel Koenig[2,4], Frédéric Lefèvre[2], Jean-Raphaël Marquès[2], Diego Oportus[2], Akihiko Ikeda[5,6], Yasuhiro H. Matsuda[5], Émeric Falize[1], and Bruno Albertazzi[2]

[1] CEA, DAM, DIF, F-91297 Arpajon, France; marin.fontaine@cea.fr
[2] LULI, CNRS, École Polytechnique, CEA, Sorbonne Universités, Institut Polytechnique de Paris, F-91128 Palaiseau cedex, France
[3] Laboratoire des Sciences du Climat et de l'Environnement, Institut Pierre Simon Laplace, CEA-CNRS-UVSQ, 91190 Gif-sur-Yvette cedex, France
[4] Graduate School of Engineering, Osaka University, Suita, Osaka 565-0871, Japan
[5] Institute for Solid State Physics, University of Tokyo, Kashiwa, Chiba 277-8581, Japan
[6] Department of Engineering Science, University of Electro-Communications, Chofu, Tokyo 182-8585, Japan

## Abstract

This study investigates the influence of Herbig–Haro jets on initiating star formation in dense environments. When molecular clouds are nearing gravitational instability, the impact of a protostellar jet could provide the impetus needed to catalyze star formation. A high-energy-density experiment was carried out at the LULI2000 laser facility, where a supersonic jet generated by a nanosecond laser was used to compress a foam or plastic ball, mimicking the interaction of a Herbig–Haro jet with a molecular cloud. Simulations using the 3D radiation hydrodynamics code TROLL provided comprehensive data for analyzing ball compression and calculating jet characteristics. After applying scaling laws, similarities between stellar and experimental jets were explored. Diagnostic simulations—including density gradient, emission, and X-ray radiographies—showed strong agreement with experimental data. The results of the experiment, supported by simulations, demonstrate that the impact of a protostellar jet on a molecular cloud could reduce the Bonnor–Ebert mass by approximately 9%, thereby initiating collapse.

*Unified Astronomy Thesaurus concepts:* Herbig-Haro objects (722); Laboratory astrophysics (2004); Hydrodynamics (1963); Stellar jets (1607); Star formation (1569)

## 1. Introduction

When a young star forms within a dense molecular cloud, an accretion disk develops that draws in surrounding matter. In response, a supersonic stellar jet is produced that can stretch across several parsecs and reach speeds around $100\,\mathrm{km\,s^{-1}}$ (J. Bally 2016). These jets, known as Herbig–Haro (HH) jets, represent a fascinating and central subject in the investigation of star formation processes. Recently, high-resolution infrared observations from the James Webb Space Telescope have offered new insights into the morphology, dynamics, and interactions of these jets with their surrounding environment (E. Habart et al. 2024). In certain cases, these jets may come into contact with other dense objects or neighboring molecular clouds, as observed in the case of HH 110 (P. Hartigan et al. 2009).

Gas clusters can collapse and transform into stars through various processes, including gravitational collapse (C. F. McKee & E. C. Ostriker 2007), magnetic fields (P. Padoan & Å. Nordlund 2011), turbulence (C. Federrath et al. 2011), and radiative condensation (C. de Boisanger & J. P. Chieze 1991). In giant gas clouds, such as the Orion Nebula (J. Bally 2016), supersonic jet ejections may contribute to the formation of new generations of stars. This illustrates how external forces, specifically HH stellar jets, can trigger star formation in molecular clouds approaching gravitational collapse. In IC 1396A, known as the Elephant Trunk nebula, external triggering has been estimated to account for 14%–25% of the total stellar formation rate (K. V. Getman et al. 2012). Jet-induced star formation has also been investigated numerically, particularly in the case of active galactic nucleus jets colliding with intergalactic clouds (P. C. Fragile et al. 2017).

Recent advancements in experimental techniques have enabled researchers to replicate astrophysical phenomena in laboratory environments using pulsed power facilities. These advancements include the application of high-energy-density laser systems (B. A. Remington et al. 2006) as well as alternative approaches, such as Z-pinch facilities (S. V. Lebedev et al. 2002, 2005). This has allowed scaled experiments of various astrophysical objects, including protostellar jets (B. Loupias et al. 2007). Laboratory studies have explored jets in dense media (J. M. Foster et al. 2005), the formation of clumpy jets via Z-pinch techniques (S. V. Lebedev et al. 2011), jet collimation through poloidal magnetic fields (B. Albertazzi et al. 2014), and the behavior of deflected jets upon impact with dense clumps (P. Hartigan et al. 2009). Additionally, external triggering has been experimentally studied in the context of supernova remnants (B. Albertazzi et al. 2022).

This paper focuses on the role of HH jets in triggering star formation within dense environments, such as molecular clouds. It presents the results of a laboratory astrophysics experiment conducted at the LULI2000 laser facility, investigating how a jet interacts with and compresses a spherical object and examining the resulting shock dynamics. In addition, this experiment was simulated using the 3D radiation hydrodynamics arbitrary Lagrange–Euler (ALE) code TROLL (E. Lefebvre et al. 2018), offering valuable complementary insights into the experimental process. The simulation provided access to physical parameters that could not be measured with

**Table 1**
Physical Properties of HH Jets (HH 110 and HH 248) and Experimental Jet Parameters Presented at Laboratory Scale (Lab. Jet)

| Parameters | HH 110 | HH 248 | Lab. Jet |
|---|---|---|---|
| Length $L$ (cm) | $1 \times 10^{15}$ | $1.5 \times 10^{17}$ | 0.2 |
| Density $\rho$ (g cm$^{-3}$) | $2 \times 10^{-20}$ | $1 \times 10^{-21}$ | $1.5 \times 10^{-2}$ |
| Pressure $P$ (dyn cm$^{-2}$) | $1 \times 10^{-8}$ | $1 \times 10^{-8}$ | $6.6 \times 10^{8}$ |
| Timescale $t$ (s) | $1 \times 10^{9}$ | $1.1 \times 10^{9}$ | $2.5 \times 10^{-8}$ |
| Velocity $V$ (km s$^{-1}$) | 150 | 700 | 50 |
| Composition | H | H | Ti |
| Temperature $T$ (eV) | 0.6 | 1–2 | 2.2 |
| Euler number Eu | 22 | 20 | 24 |
| Mach number $M$ | 10 | 40–55 | 6 |
| Density ratio $\eta$ | 10–100 | 10 | 70 |
| Aspect ratio $\theta$ | ∼10 | ∼10 | 3.1 |
| Cooling parameter $\chi_{\rm jet}$ | >1 | 1.3–2 | 1.4 |
| Cooling parameter $\chi_{\rm cloud}$ | ≫1 | ≫1 | ∼10 |

**Note.** The HH 110 and HH 248 jet values are, respectively, taken from P. Hartigan et al. (2009) and G. Y. Liang et al. (2018).

the diagnostics and allowed for the study of the jet and cloud behavior at later times.

The following Section 2 explains the scaling between astrophysical and laboratory scales. Section 3 details the experimental setup, followed by Section 4, which describes the TROLL and DIANE codes. Section 5 highlights the alignment between simulations and experimental results. Last, Section 6 discusses the capacity of protostellar jets to compress molecular clouds and induce collapse, supported by both experimental and simulation findings, and offers perspectives on future research.

## 2. Similarity Properties

To replicate the interaction between a protostellar jet and a molecular cloud in the laboratory, it is essential to explore the scaling laws that bridge these systems. Rather than focusing on the mechanisms of jet formation, our objective is to reproduce the hydrodynamic characteristics of jet propagation, closely resembling those observed on astrophysical scales.

Scaling laws are fundamental for connecting astrophysical and laboratory jet systems, as they must adhere to the same underlying physics (D. Ryutov et al. 1999; É. Falize et al. 2011). Table 1 details the parameters of the laboratory jet used in this experiment in comparison to two HH jets. The cooling parameter, $\chi_{\rm jet}$, which represents the ratio of the cooling time to the jet's dynamical time, is approximately 1, indicating that radiative effects significantly influence jet dynamics (J. M. Blondin et al. 1989). However, for the shock transmitted to the molecular cloud, the cooling parameter $\chi_{\rm cloud}$[7] is greater than 1, suggesting that radiative effects are minimal in the context of cloud interaction and compression. Both the laboratory and astrophysical jets consist of plasmas that behave as nonviscous and compressible flows. Thus, they can be accurately described by the Euler equations (e.g., Y. B. Zel'dovich & Y. P. Raizer 1967; L. D. Landau & E. M. Lifshitz 1987), with radiative effects omitted in this context ($\chi \gg 1$), such as:

$$\frac{\partial \rho}{\partial t} + \boldsymbol{\nabla} \cdot (\rho \boldsymbol{v}) = 0, \quad (1)$$

$$\rho \frac{d\boldsymbol{v}}{dt} + \boldsymbol{\nabla} P = 0, \quad (2)$$

$$\frac{\partial}{\partial t}\left(\rho \epsilon + \frac{\rho v^2}{2}\right) + \boldsymbol{\nabla} \cdot \left[\rho \boldsymbol{v}\left(\epsilon + \frac{v^2}{2}\right) + P\boldsymbol{v}\right] = 0, \quad (3)$$

where $\rho$, $\boldsymbol{v}$, $P$, and $e$ are, respectively, the density, velocity, pressure, and internal energy. The fluid is assumed to behave as a polytropic flow, meaning its internal energy is directly proportional to its pressure (D. D. Ryutov & B. A. Remington 2003), as described by the following relationship, with $\gamma$ being the adiabatic index:

$$\rho e = \frac{P}{\gamma - 1}. \quad (4)$$

These equations remain invariant under the rescaling (D. Ryutov et al. 1999):

$$r' = ar, \quad \rho' = b\rho, \quad P' = cP, \quad (5)$$

where $a$, $b$, and $c$ are the constant scaling factors, providing the following constraints:

$$t' = a\sqrt{\frac{b}{c}}\, t, \quad v' = \sqrt{\frac{c}{b}}\, v. \quad (6)$$

The values summarized in Table 1 allow for the calculation of constant scaling factors for both the HH 110 and HH 248 jets. These calculations follow Equations (5), where each parameter $x$ corresponds to the astrophysical scale and $x'$ to the laboratory scale.

For HH 110, with scaling parameters $a \sim 2 \times 10^{-16}$, $b \sim 7.7 \times 10^{17}$, and $c \sim 6.7 \times 10^{16}$, 2 yr (or $5.9 \times 10^{7}$ s) for an HH jet corresponds to approximately 40 ns in the laboratory and an astrophysical jet velocity of 170 km s$^{-1}$ corresponds to 50 km s$^{-1}$ for a laboratory jet.

For HH 248, using $a \sim 1.3 \times 10^{-18}$, $b \sim 1 \times 10^{21}$, and $c \sim 6.7 \times 10^{16}$, the same scaling translates 64 yr (or $2 \times 10^{9}$ s) for an HH jet to about 40 ns in the laboratory, with a 750 km s$^{-1}$ astrophysical velocity corresponding to 50 km s$^{-1}$.

The scaling also preserves the dimensionless Euler number $Eu = v\sqrt{\rho/p}$, which is greater than 1 for all jets. Additionally, the jets exhibit a high internal Mach number $M$, defined as the ratio of the jet velocity to the internal sound speed, indicating supersonic speeds and similar physical regimes.

To further match HH jets, two dimensionless numbers are essential. The first is the aspect ratio, $\theta$, representing the jet's length-to-width ratio. This aspect ratio, shown in Table 1, ensures that the jet's shape upon impact with the ball resembles that observed in astrophysical systems. In this work, the small cloud approximation was made, where the cloud diameter is comparable to the jet width (R. I. Klein et al. 1994). Therefore, the jet's pointed shape, rather than its entire length, governs the collision dynamics. The second parameter is the density ratio between the ball and the jet, $\eta = \rho_{\rm cloud}/\rho_{\rm jet}$. The ball density in the laboratory is set to approximate the density ratio between an HH jet and a molecular cloud. This parameter significantly affects the velocity of the shock transmitted into the ball and consequently its compression by the jet.

---
[7] Here, the shock's crossing time over the cloud is taken as the dynamical time.

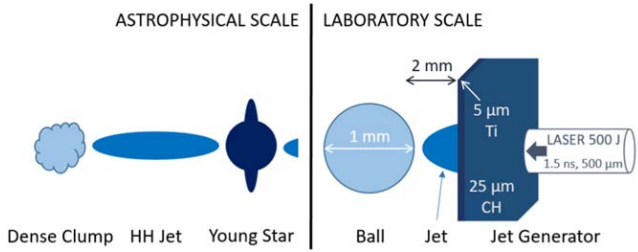

**Figure 1.** Astrophysical configuration compared to the experimental setup.

These scaling parameters and dimensionless numbers highlight the similarity between laboratory and astrophysical scales.

### 3. Experimental Setup

The experiment was conducted at the LULI2000 laser facility (M. Koenig et al. 2006). The setup (shown in Figure 1) involved a laser pulse of duration $\tau \sim 1.5$ ns and energy $E_0$ between 400 and 600 J, at a wavelength of $\lambda_L = 527$ nm. This pulse irradiated a jet generator, focused on a 500 $\mu$m diameter flat-topped spot using a hybrid phase plate, achieving an intensity of $I_{\text{laser}} \sim 1 \times 10^{14}$ W cm$^{-2}$.

The jet generator was designed with a two-layer target (see Figure 1). The first layer, a 25 $\mu$m thick plastic (CH) sheet, prevented laser reflection and efficiently converted laser energy into kinetic energy, initiating a shock through the material via the rocket effect. This shock then propagated into a second layer, a 5 $\mu$m thick sheet of titanium, which acted as a shield by absorbing hard X-rays produced during the laser–plasma interaction, thereby preventing preheating of the obstacle. As the shocked material expanded into the vacuum, it formed the desired jet. Figure 2 illustrates this sequence, detailing the laser–target interaction and showing the jet propagating downward and impacting the ball. The obstacle, positioned 2 mm from the jet generator (measured from the edge of the target to the ball's edge) and aligned with the focal spot, was either a plastic ball ($C_8H_8$ with $\rho = 1.05$ g cm$^{-3}$) or a foam ball (CHO with $\rho = 60$ mg cm$^{-3}$), with 1 mm diameter. This placement ensured optimal jet formation and facilitated the study of its morphology and behavior in the vacuum.

Various optical diagnostics were employed to characterize the system. Self-emission from 580 to 750 nm was captured via 1D time-resolved and 2D spatially resolved optical pyrometry, tracking the jet's velocity and morphology before impact. Additionally, time-resolved (1D) and space-resolved (2D) shadowgraphy, performed with a low-energy probe beam (a few mJ, ~10 ns pulse duration), allowed boundary measurements at regions of refraction, generally below the critical density ($n_e \lesssim 1 \times 10^{20}$ cm$^{-3}$). This shadowgraphy revealed the jet's contours, velocity, and structure prior to impact.

Finally, down–up X-ray radiography provided high-resolution imaging of the shock transmitted into the obstacle. A 25 $\mu$m diameter titanium wire, irradiated with a 400 J laser pulse of 0.5 ns duration at 527 nm, focused to a spot slightly larger than the wire diameter, produced the X-ray source for this diagnostic. Positioned 2.5 cm below the obstacle, this wire emitted X-rays captured by an image plate located 60 cm above, with a geometrical magnification of approximately 24. Filtering with 100 $\mu$m of polyethylene and 11.4 $\mu$m of titanium, forming a Ross filter pair, created a quasi-monochromatic

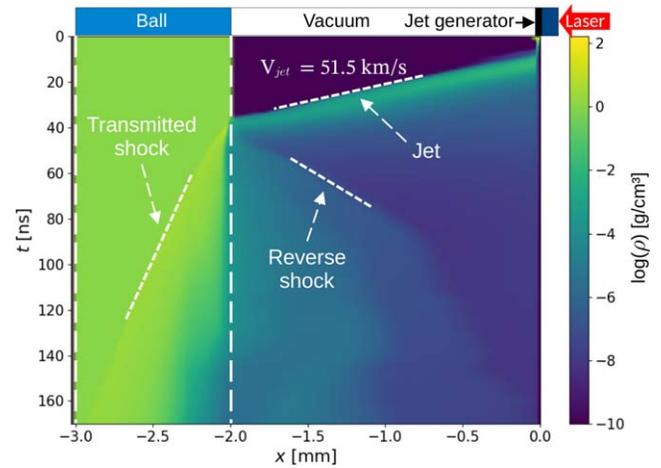

**Figure 2.** Simulated spacetime diagram of the density along the symmetry axis. The laser impacts from the right, initiating jet propagation in vacuum from right to left and downward. Upon contacting the ball, both forward and reverse shocks form, compressing the ball.

X-ray radiograph at the He-$\alpha$ line (~4.7 keV), reducing bremsstrahlung contributions from the coronal plasma. This monochromatic X-ray setup allowed precise density measurements using Beer–Lambert's law. The spatial resolution of the X-ray radiography, mainly constrained by source size and shock motion blur, achieved an effective resolution of around 30 $\mu$m. Although the 500 ps pulse duration led to an expected 2.5 $\mu$m blurring (due to the shock velocity of ~5 km s$^{-1}$), this effect was minimal compared to the dominant 25 $\mu$m blurring due to the X-ray source size itself.

In this setup, a solid sphere serves as the laboratory analog of a small molecular cloud. Under the small-cloud approximation, the supersonic transmitted shock propagates through the cloud faster than pressure adjustments can occur within it. The two critical parameters governing this interaction are the density ratio $\eta$ and the velocity of the incoming shock. These quantities remain consistent between the laboratory and astrophysical scales, ensuring the validity of the comparison, as described in previous studies (e.g., P. Hartigan et al. 2009; B. Albertazzi et al. 2022).

A total of nine shots were conducted with an unvarying experimental setup. The laser energy was the only variable, ranging from 400 to 600 J. In one specific shot, the ball was removed, to enable a detailed investigation of the jet's behavior at later stages.

### 4. Numerical Setup

All simulations were conducted using the Commissariat à l'Énergie Atomique et aux Énergies Alternatives (CEA)'s 3D radiation hydrodynamics code, TROLL (E. Lefebvre et al. 2018). This ALE multimaterial code is designed to compute laser-driven target dynamics at laboratory scales. TROLL solves the equations for mass, momentum, and energy conservation, while utilizing a multigroup implicit Monte Carlo approach for radiation transport—in this case, approximately 200 groups. The laser beam is discretized into numerical rays, which are traced through the plasma to account for propagation, refraction, and absorption. Material properties, such as the equation of state, opacity, and emissivity, are calculated in local thermodynamic equilibrium using specialized codes and then tabulated. TROLL can thus simulate an entire experiment, from the laser pulse to the propagation of the supersonic jet and compression of the target ball. The numerical configuration closely mirrored the experimental setup. The simulation box consisted of a fixed cylindrical tube with a 4 mm diameter and open ends. The laser beam, discretized into numerical rays, interacted with a bimaterial target composed of CH and Ti sheets. The resulting jet expanded into a vacuum and impacted the ball positioned 2 mm from the target. The simulation was executed on 512 processors, involving approximately $2 \times 10^6$ cells and requiring 150,000 CPU hours to complete. Each cell in the jet generator plate measured approximately 1 $\mu$m in thickness. A mesh convergence study was conducted to evaluate the influence of the cell thickness on the laser absorption and the characteristics of the generated jet. The simulation accurately captured the entire sequence of events, including the laser–target interaction, jet formation, impact on the ball, subsequent compression, and the jet's evolution over a period of approximately 200 ns.

To refine experimental comparisons, additional X-ray radiography post-process simulations were conducted with the DIANE code (M. Caillaud et al. 2014). DIANE uses ray tracing to simulate how an X-ray source spectrum interacts with the target, accounting for its geometry and hydrodynamic state. For these X-ray simulations, a monochromatic spectrum was chosen at the titanium resonance energy of ∼4.7 keV, enhancing the accuracy of the results observed on the detector.

## 5. Results and Numerical Study

The experiment involved a total of nine shots, with general parameters summarized in Table 2. One of these shots was conducted without a ball, to examine the jet's dynamics and behavior over an extended duration.

### 5.1. Simulation of the Experiment

The 3D simulation of the experiment, using the same setup as the actual experimental configuration, is shown at three distinct time points (30, 60, and 90 ns) in Figure 3. With the jet velocity as the primary matching parameter, the laser energy in the simulation was reduced, to achieve a numerical jet velocity of approximately 50 km s$^{-1}$. The snapshots in Figure 3 illustrate the jet's morphology at different stages: prior to impact (left panel), during impact with the ball and subsequent shock transmission (middle panel), and as the jet envelops the ball, propagating the shock farther (right panel). As the jet envelops the ball, it undergoes a noticeable shape change, expanding with a larger radius. This effect is significant for drawing comparisons with astrophysical phenomena, where aspects like the jet's deflection angle or its enlargement could be compared to astronomical observations. Such observations could help constrain parameters like the jet's aspect ratio or the ball's size. However, this phenomenon is observable only through simulation, as it requires shadowgraphy diagnostics (low-emission and low-density) that were not performed post-collision in the experiment.

The agreement between the experimental and simulated results is supported by qualitative and quantitative analyses, detailed in this section.

**Table 2**
Experiment Shots Summary

|  | Jet Only | Foam Ball | Plastic Ball |
|---|---|---|---|
| Number of shots | 1 | 3 | 5 |
| Energy (J) | 621 | 524 | 525 |
| Velocity (km s$^{-1}$) | 53.4 ± 1.3 | 50.5 ± 3.0 | 53.8 ± 3.1 |
| Impact time (ns) | ⋯ | 40.6 ± 0.1 | 38.2 ± 0.1 |

**Note.** The first column lists the number of shots for each configuration, while the remaining columns provide averaged measurements across these shots. The impact time was calculated based on the jet's velocity, which remained constant over the distance to the ball.

### 5.2. The Supersonic Jet

To understand the ball's compression process, we must first analyze the characteristics of the supersonic jet before impact. Key properties of the jet—such as its velocity and impact time—were measured, and these results are provided in Table 2. Additional detailed characteristics, including data from both experimental measurements and simulations, are shown in Table 1.

The variations in laser energy across the shots led to heterogeneous jet propagation velocities. The 1D shadowgraphy diagnostic, shown in the top panel of Figure 4, was used to determine the propagation velocity in the vacuum separating the jet generator and the ball. The jet propagates from the top right to the bottom left, with its velocity marked by a white line, while the ball's outline is shown by dashed white lines. The experimental jet velocities, ranging from 50 to 54 km s$^{-1}$, along with their measurement precision, are presented in Table 2, while the simulated jet velocity is approximately 52 km s$^{-1}$. This precision accounts for laser energy variability, the accuracy of the 1D shadowgraphy diagnostic, and the post-processing method used for velocity determination. The velocity distribution is depicted in Figure 5, with each point representing a shot and its associated precision. Three points marked in red correspond to the shots whose X-ray diagnostics are used to analyze the ball's compression.

The jet's morphology and scale are visualized through the 2D shadowgraphy diagnostic in the bottom panel of Figure 4. This diagnostic was conducted early in the experiment, to capture the jet's morphology pre-collision, highlighting its collimated shape with an aspect ratio of $\theta \sim 3$. The contour of the simulated jet is also represented in this figure by the red dotted line, revealing that the same jet morphology and aspect ratio are obtained. More precisely, the impact and resulting compression are governed by the pointed tip of the jet rather than its full length. This setup is thus well suited to describing the interaction of protostellar jets with larger aspect ratios in the small-cloud approximation.

These findings confirm the relevance of the 3D simulations, demonstrating that they accurately replicate the jet's behavior as observed experimentally. Consequently, we can confidently use the simulations to derive additional hydrodynamic properties of the jet, such as temperature and pressure, which are challenging to measure directly in the experiment. These computed properties are summarized in Table 1.

### 5.3. Study of the Ball's Compression

The supersonic jet moving through the vacuum impacts the molecular cloud, represented by a ball in the experiment,

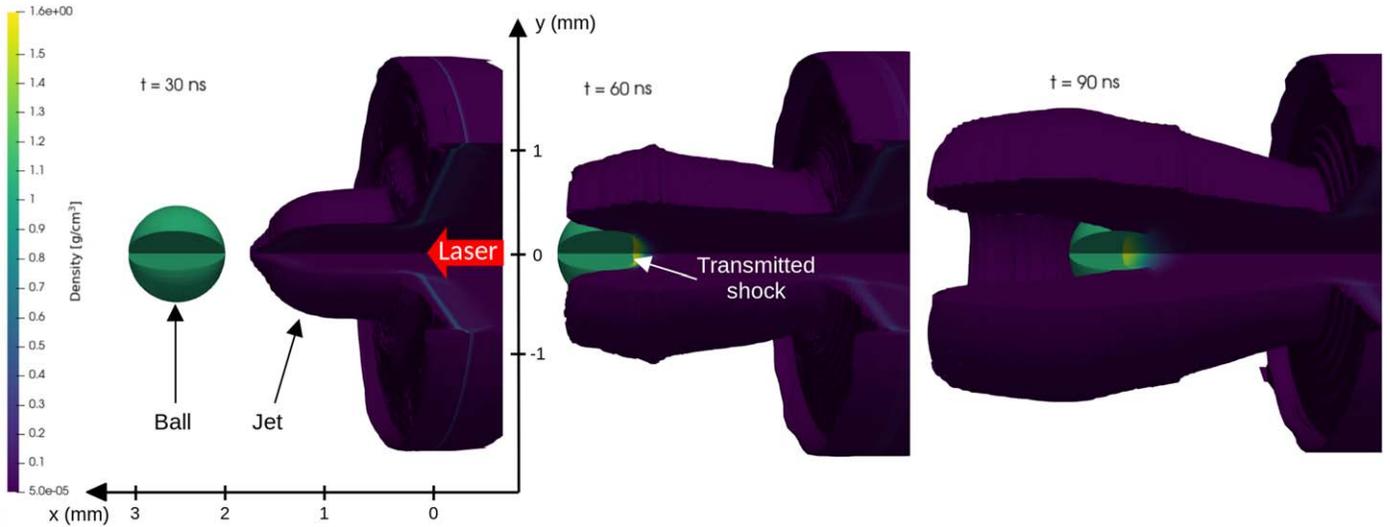

**Figure 3.** Logarithmic density map of the simulated 3D jet at three different times. The laser is coming from the right and indicated with the red arrow. A section of the 3D simulation has been removed to reveal the interior structure, particularly highlighting the compression of the ball.

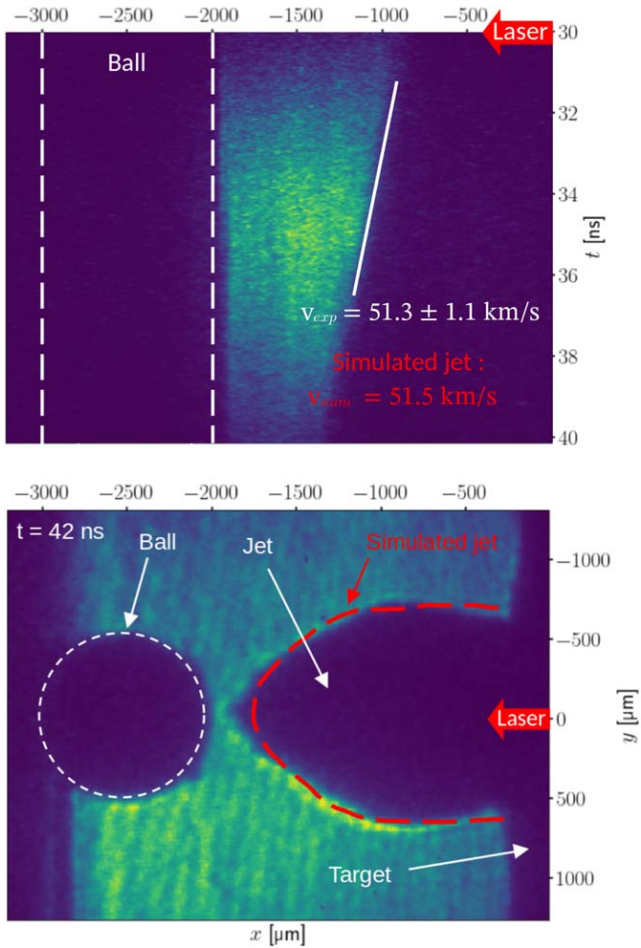

**Figure 4.** Annotated 1D (top) and 2D (bottom) shadowgraphy images of the jet prior to collision. The simulated jet's position and contour at the same time are shown with the red dashed lines.

triggering a weak shock that propagates within it. X-ray radiographs of the ball were captured at specific intervals, between 100 and 200 ns, allowing for observation of the shock's progression through the ball, from which its shape and velocity were measured.

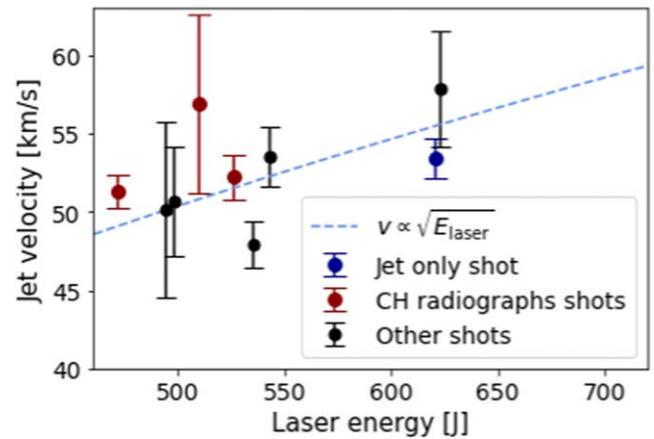

**Figure 5.** Velocity distribution of the shots (black dots) in relation to the laser's energy. Shots involving plastic balls, used for three radiography diagnostics, are indicated by red dots, while the blue dot corresponds to the jet-only shot.

In the first part of this study, we examine shots involving a plastic ball ($C_8H_8$, $\rho = 1.05$ g cm$^{-3}$). Radiographs from three different shots taken at 100, 160, and 200 ns are shown in Figure 6, with experimental radiography in the top half and simulated radiography below. The ball and shock boundaries are marked with white lines for clarity. Due to the shot-to-shot variations in jet velocity, as depicted in Figure 5, a time shift must be considered in the simulation. For instance, two highlighted shots, performed at the same laser energy, show a 5 km s$^{-1}$ difference in jet velocity. Although both jets reach the same velocity, the simulated jet exits the generator sooner, resulting in an earlier impact and faster shock initiation. Despite the timing difference, there is strong qualitative agreement between the experimental and simulated X-ray radiographs. Both the shape and contrast of the shock align well, though the simulation shows a slightly faster transmitted shock. This minor lead, however, does not impact the overall compression process. During compression, the density behind the shock in the plastic ball reaches approximately 1.5 g cm$^{-3}$, with a final compression factor $C_f = \rho_f/\rho_0 = 1.20$—values that the simulation reproduces accurately.

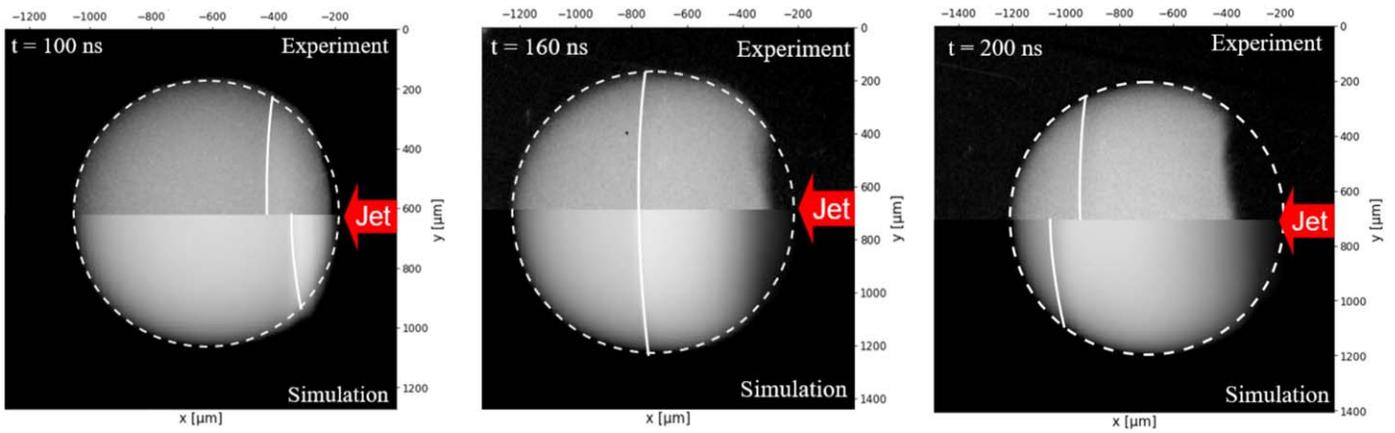

**Figure 6.** X-ray radiographs from three different shots with plastic balls at various times. Each image shows the experimental X-ray radiography in the top half and the corresponding simulated radiography below. The jet, indicated by a red arrow, impacts the ball along its axis. The shock position and initial ball outline are highlighted by the solid and dashed lines, respectively, for clarity.

The compression study was also extended to foam balls (CHO, $\rho = 60$ mg cm$^{-3}$). Radiographs at 130 and 160 ns are presented in Figure 7, where time shifts between experimental and numerical results are again taken into account. At 130 ns, the simulated shock and ball shape align well with experimental observations. However, by 160 ns, the simulated jet has already fully disrupted the ball, due to the faster shock propagation observed. This disruption in the foam ball is attributed to a lower density ratio ($\eta \approx 4$, compared to the previous value of 70). As a result, the foam ball cannot withstand the jet's impact and is blown away in both the second radiograph and at later times in the simulation. These findings suggest a minimum density ratio threshold is needed for the ball to sustain the shock's passage without complete disintegration.

The ball's density profile along its axis, assuming symmetry and a monochromatic source, can be derived from X-ray radiographies using Beer–Lambert's law. The intensity of the X-ray beam after passing through the ball $I_l$ can be expressed as

$$I_l = I_0 \exp(-\kappa \rho d), \quad (7)$$

where $I_0$, $\kappa$, and $d$ are, respectively, the intensity of the X-ray source, the mass attenuation coefficient, and the distance the rays travel through the ball.

For the plastic ball, density profiles at 100 and 200 ns are shown in Figure 8. Initially, the shock's compression factor is approximately 1.5, as seen in the lighter blue curve at 100 ns, with the shock position marked by a red dashed line. By 200 ns (darker blue curve), the shock becomes less distinct but is estimated near $x \simeq -0.7$ mm in the radiograph, showing that the shock has densified the ball to an average density of approximately $\rho_f \approx 1.25$ g cm$^{-3}$, estimated using Equation (7). The compression also reduces the ball's size by about 10%, suggesting a conservation of mass, while the foam ball may experience greater mass loss due to lower density and greater jet disruption (as seen in Figure 7).

Theoretically, the jet impact can be considered as a shock wave interacting with the molecular cloud. The impact results in the generation of a weak reflected shock and a slower transmitted shock propagating into the cloud (C. F. McKee & L. L. Cowie 1975). Assuming the shock is nonradiative, one can derive the velocity of the transmitted shock into the ball as

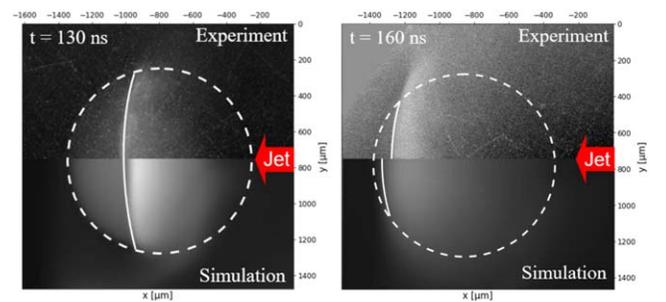

**Figure 7.** X-ray radiographs from two different shots with foam balls at various times. Each image shows the experimental X-ray radiography in the top half and the corresponding simulated radiography below. The jet, indicated by a red arrow, impacts the ball along its axis. The shock position and initial ball outline are highlighted by solid and dashed lines, respectively, for clarity.

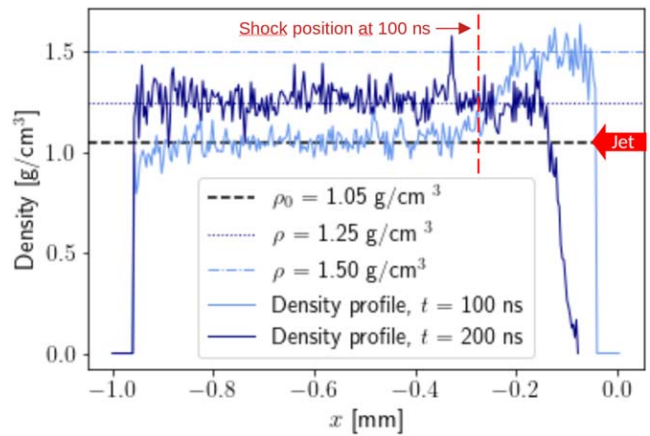

**Figure 8.** Density profiles of the plastic ball at 100 and 200 ns, obtained from X-ray radiographies of two separate shots.

follows (E. P. G. Johansson & U. Ziegler 2013):

$$v_s \simeq \frac{v_j}{\eta^{1/2}}. \quad (8)$$

Here, $v_s$, $v_j$, and $\eta$ represent the velocity of the transmitted shock into the cloud, the jet's velocity, and the density ratio, respectively, as defined earlier and presented in Table 1. The theoretical transmitted shock velocities for the plastic and foam balls were calculated as $v_{s,th} \simeq 6$ km s$^{-1}$ and $v_{s,th} \simeq 35$ km s$^{-1}$, respectively, using the density ratios $\eta = 70$ and $\eta = 4$. In the case of the plastic ball, as depicted in Figure 9, the experimental shock velocity $v_s \approx 4.2 \text{ km s}^{-1}$ (blue dashed line) can thus be compared with the velocity obtained in the simulation $v_{s,\text{num}} \approx 5.2 \text{ km s}^{-1}$ (green dashed line) and the theoretical prediction $v_{s,\text{th}} \approx 6.0 \text{ km s}^{-1}$ (red dashed line).

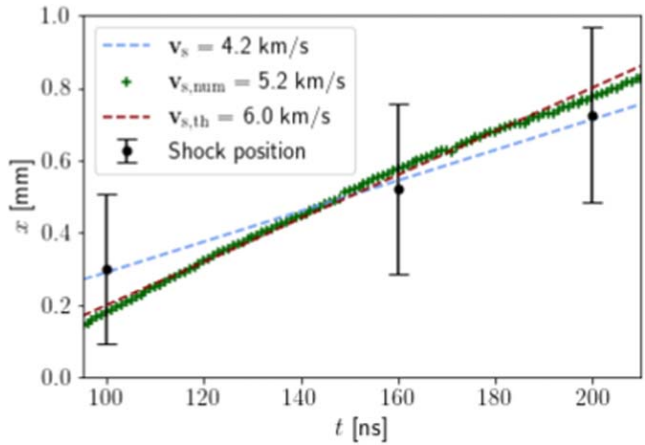

**Figure 9.** The evolution of the shock position in the plastic ball (black dots and blue dashed line) compared to the shock velocity obtained from the simulation (green crosses) and the theoretical velocity (red dashed line).

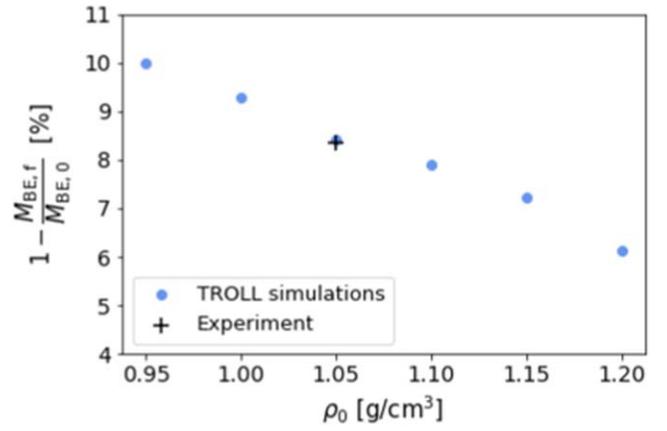

**Figure 10.** Evolution of the reduction rate of the Bonnor–Ebert mass following interaction with the jet. The experimental result (black cross) is compared with multiple 3D simulations conducted at various initial densities of the plastic ball (blue dots).

The experimental study on ball compression, corroborated by the simulation, provides insight into shock propagation mechanics and density effects, demonstrating effective agreement in shock shape, density profiles, and velocity trends across plastic and foam materials.

## 6. Discussion and Perspectives

The diagnostics from this experiment provide crucial insights into the behavior of transmitted shocks within molecular clouds, particularly in the context of star formation and the interactions between HH jets and dense clumps. As observed in Section 5, when HH jets collide with these dense clumps, they can enhance the density and alter the physical characteristics of the cloud, potentially leading to gravitational collapse. To quantify the gravitational stability of such a dense clump, the Bonnor–Ebert mass $M_{\text{BE}}$ before interaction with an HH jet needs to be calculated. This mass indicates the critical point at which the internal pressure forces of a gas cluster are no longer sufficient to oppose gravitational forces. This mass is expressed as (C. F. McKee & E. C. Ostriker 2007)

$$M_{\text{BE}} = 1.182 \frac{\sigma_{\text{th}}^3}{(G^3 \rho_0)^{1/2}}, \quad (9)$$

where $P_{\text{th},0} = \sigma_{\text{th}}^2 \rho_0$ is the thermal pressure, $\sigma_{\text{th}}$ is the sound speed, $\rho_0$ is the density, and $G$ is the gravitational constant. From the radiographies depicted in Figure 6, we see that the ball's size decreases after interaction with the jet, while the shock wave increases the mean density and pressure within the ball. Despite these changes, it is reasonable to assume that the thermal pressure $P_{\text{th},0} = \sigma_{\text{th}}^2 \rho_0$ remains constant before and after the impact, as the weak shock ensures that the sound speed is kept unchanged throughout the process. This consistency has been corroborated by simulation results. By introducing the compression factor $C_f = \rho_f / \rho_0$, the Bonnor–Ebert mass post-densification can be expressed as

$$M_{\text{BE,f}} = \frac{M_{\text{BE,0}}}{C_f^{1/2}}. \quad (10)$$

For the plastic ball, a compression factor of $C_f \sim 1.20$ was determined (see Figure 8), resulting in a reduction of the Bonnor–Ebert mass by approximately 9%. This aligns with the observed volume reduction of around 10%, suggesting that the cloud retains its mass throughout the interaction with the jet. The implications for molecular clouds near the Bonnor–Ebert mass—essentially on the brink of gravitational instability—are significant; the energy imparted by the jet could potentially induce collapse. The evolution of the Bonnor–Ebert mass post-interaction is illustrated in Figure 10, where the experimental result is compared with various 3D simulations across different initial densities of the plastic ball. The simulations span a range of initial densities from $0.95 \text{ g cm}^{-3}$ to $1.20 \text{ g cm}^{-3}$, corresponding to density ratios $\eta$ between 63 and 80. The analysis indicates that lower initial densities yield higher compression factors, provided the cloud withstands the interaction.

In contrast, the foam ball presents a different outcome. As previously noted, a definitive final compression factor cannot be established, as the jet completely disrupts its structure. This outcome emphasizes the necessity for an adequate density ratio between the jet and the molecular cloud. If the cloud lacks sufficient density, it cannot maintain its integrity under impact, resulting in its destruction and negating any possibility for collapse and subsequent star formation. This required minimum density ratio must exceed 4, although this specific threshold was not directly investigated in this study.

The dynamics of molecular clouds and their interactions with external forces are critical in understanding star formation processes. J. P. Chieze (1987) illustrated that isothermal clouds adhere to a mass–radius relationship characterized by $M \propto R^2$, establishing a framework for discerning stable and unstable states based on a critical point on a log–log graph of mass versus radius. This framework has been applied to a population of approximately 100 molecular clouds, indicating that some are nearing gravitational instability. In particular, a study of the Vela molecular cloud through a BLAST survey examined 141 prestellar and protostellar dust cores (L. Olmi et al. 2010), revealing that around 10 of these cores are in close proximity to instability when comparing the core mass to the Bonnor–Ebert mass. These findings suggest that a significant number of clouds could transition into a collapsing regime, due to interactions with HH jets, even if the initial momentum from the jet is insufficient. The propagation of compression shocks within the cloud, coupled with a reduction in stability criteria, has the potential to initiate collapse. Furthermore, other physical phenomena, beyond gravity, can play a pivotal role in enhancing collapse dynamics. For instance, radiation processes can amplify and accelerate the collapse of molecular clouds (C. de Boisanger & J. P. Chieze 1991). Turbulence generated by shocks traversing nonuniform media, such as fragmented clouds, also contributes to this process. Research indicates that turbulence driven by such shocks may lead to higher star formation rates compared to solenoidal driving mechanisms (S. Dhawalikar et al. 2022). The interplay of turbulence and gravitational effects can further elevate the rate of star formation (E. Jaupart & G. Chabrier 2020).

An illustrative case is the investigation of IC 1396A, or the Elephant Trunk nebula, where observations from Chandra have revealed evidence of triggered star formation at the edges of the H II region. This study identified more than 140 stars, representing 14%–25% of the total population, thereby supporting the concept of external triggering in star-forming regions (K. V. Getman et al. 2012). The findings presented in this paper enhance the theoretical and observational understanding of the mechanisms that facilitate cloud collapse and subsequent star formation.

Looking ahead, future research could investigate the interaction of jets with molecular clouds under varying configurations, such as altering the jet's aspect ratio and density or adjusting the size of the cloud relative to the jet. Such studies would provide deeper insights into the criteria necessary for triggering collapses and potential new star formation in larger molecular clouds through HH jet interactions. Additionally, the exploration of external triggering mechanisms in star formation could be expanded to include other phenomena, such as weaker shocks from supernova remnants.

## 7. Conclusion

This study successfully demonstrates the experimental recreation of an HH jet's compression effects on a dense molecular cloud, represented by a ball, while adhering to relevant scaling laws. Utilizing a supersonic jet propagating in a vacuum, the experimental setup mimics the physical characteristics of an HH jet, enabling an in-depth examination of the interaction between the jet and the cloud, as visualized through X-ray radiographies. The ability to numerically replicate the experiment and its diagnostics in 3D further supports the findings. The results indicate that a molecular cloud can endure interaction with an HH jet, provided that the density ratio is sufficiently high. The jet's impact densifies and reduces the size of the cloud, as supported by numerical investigations into the relationship between the initial density and compression factor, as well as Bonnor–Ebert mass reduction.

Crucially, this research highlights that a stable molecular cloud, close to its gravitational equilibrium mass limit, can be driven into an unstable regime upon interaction with a protostellar supersonic jet, as its critical mass is reduced by approximately 9%. Notably, it is shown that as long as the cloud maintains an adequate density ratio and does not disintegrate due to the jet, a lower initial density correlates with a higher compression factor. These insights contribute to the understanding of external triggering mechanisms, particularly those exerted by HH jets, in the star formation process. Such external influences could play a significant role in the evolution of star-forming regions, including molecular clouds and nebulae, shaping the dynamics of star birth across the cosmos.

### ORCID iDs

Marin Fontaine ● https://orcid.org/0009-0000-6974-2692
Clotilde Busschaert ● https://orcid.org/0000-0002-6975-6581
Diego Oportus ● https://orcid.org/0009-0002-3706-5716
Akihiko Ikeda ● https://orcid.org/0000-0001-7642-0042
Yasuhiro H. Matsuda ● https://orcid.org/0000-0002-7450-0791